\documentclass[shortnote,twocolumn]{jpsj2}
\usepackage{graphicx}
\usepackage{bm}

\newcommand{\irm}{{ i}}
\newcommand{\ie}{{i.e.}}                     % that is
\newcommand{\eg}{{e.g.}}                     % for example
\newcommand{\pdag}{{\phantom{\dagger}}}
\newcommand{\ext}{\textrm{ext}}

\newcommand{\DC}{\textrm{DC}}
\newcommand{\GZ}{\textrm{GZ}}
\newcommand{\classical}{\textrm{cl}}

\newcommand{\inter}{\textrm{int}}

\newcommand{\omax}{{\omega_\textrm{max}}}

\newcommand{\toh}{{\textstyle{\frac{1}{2}}}}

\newcommand{\Hartree}{\textrm{Hartree}}
\newcommand{\imp}{\textrm{imp}}
\newcommand{\tauel}{{\tau_\textrm{el}}}
\newcommand{\tauphi}{{\tau_\varphi}}

\newcommand{\vF}{{v_\textrm{F}}}

\newcommand{\dis}{\textrm{dis}}

\newcommand{\Tr}{\textrm{Tr}}

\newcommand{\Eq}[1]{Eq.~(\ref{#1})}        % Eq. (#) Equation
\newcommand{\Eqs}[1]{Eqs.~(\ref{#1})}      % Eqs. (#) Equations
\newcommand{\bari}{{\bar \imath}}
\newcommand{\barj}{{\bar \jmath}}

\newcommand{\ij}{{i \! j}}
\newcommand{\bmR}{{\mib{R}}}

\newcommand{\bmr}{{\mib{r}}}
\newcommand{\bmp}{{\mib{p}}}
\newcommand{\bmk}{{\mib{k}}}
\newcommand{\bmq}{{\mib{q}}}

\newcommand{\bmP}{{\mib{P}}}

\newcommand{\bnabla}{{\mib{\nabla}}}

\newcommand{\G}{{\tilde G}}

\newcommand{\dsr}{{\delta \bar S_R}}
\newcommand{\sic}{{\bar S_I^\classical}}
\newcommand{\src}{{\bar S_R^\classical}}
\newcommand{\bsi}{{\bar S_I}}
\newcommand{\bsr}{{\bar S_R}}

\def\runtitle{Decoherence: Influence Functional vs.\ and Diagrams}
\def\runauthor{Jan \textsc{von Delft}}

\title{%
Decoherence of interacting electrons in disordered conductors: on the
relation between influence functional and diagrammatic approaches
}

\author{%
Jan \textsc{von Delft}\thanks{vondelft@theorie.physik.uni-muenchen.de}
}

\inst{%
Lehrstuhl f\"ur Theoretische Festk\"orperphysik, CeNS und Sektion
Physik, LMU M\"unchen, Theresienstr. 37, D-80333 M\"unchen
}

\recdate{August 7, 2002}

\abst {We establish a connection between the influence functional
  approach of Golubev and Zaikin (GZ) and  Keldysh diagrammatic
  perturbation theory for calculating the decoherence time $\tauphi$
  of interacting electrons in disordered  metals; we show how the
  standard diagrams for the Cooperon self energy can be recovered from
  GZ's influence functional $e^{- (i \bsr +  \bsi)}$. This allows us
  to shed light on GZ's claim that $\bsr$ is irrelevant for
  decoherence: $\bsr$ generates as many important self energy diagrams
  as $\bsi$;  GZ's neglect of $\bsr$ is permissible  only at high
  temperatures ($T \gtrsim \hbar / \tauel$).  }
\kword{%
Decoherence, influence functional, diagrammatic
perturbation theory, interactions, disorder
}

\begin{document}
\sloppy
\maketitle

\section{Introduction}

%One of the most fruitful techniques available for analyzing the
%decoherence properties of quantum systems coupled to  an effective
%environment is the Feynman-Vernon influence functional \cite{FV}:
%starting from a functional integral that describes the coupled
%dynamics of the quantum system plus environment, the influence
%functional is obtained by integrating out  the environmental degrees
%of freedom, and it describes, in terms of a (typically dissipative)
%contribution to the  effective action for the quantum system, how the
%dynamics of the latter is affected by the environment.  
%A famous
%example of this approach is the  Caldeira-Leggett treatment of quantum
%Brownian motion (free particle or harmonic oscillator coupled
%to bosonic heat bath), or the spin-boson problem (spin coupled to a
%bosonic heat bath). 

A few years ago, Golubev and Zaikin (GZ) developed an  influence
functional approach for describing  interacting fermions in a
disordered conductor. Their key idea was as follows: to understand how
the diffusive behavior of a given electron is affected by its
interactions with other electrons in the system, which constitute its
effective environment,  the latter should be  integrated out, leading
to an influence functional (denoted by $e^{-(i \bar S_R + \bar S_I)}$
below) in the path integral $\int {\cal D} (\bmR \bmP)$ describing its
dynamics. To derive the effective action  $i \bar S_R + \bar S_I$,  GZ
devised a strategy which, when implemented with sufficient care,
\emph{properly incorporates the Pauli principle} -- this is essential,
since both the particle and its environment  originate from the same
system of indistinghuishable fermions, a feature which  makes the
present problem interesting and sets it apart from all other
applications of the influence functional strategy that we are aware of.

GZ used their new approach  to calculate the electron 
decoherence time  $\tauphi (T)$, as extracted from
the magnetoconductance in the weak localization regime,
and found it to be finite at zero temperature:\cite{GZ1,GZ2,GZ3,GZS}
%\begin{eqnarray}
%  \label{eq:GZmainclaim}
$\tauphi (T \to 0) = \tau_\varphi^0$,
%\end{eqnarray}
%$\gamma_\phi (0) \neq 0$,
in apparent  agreement with some experiments \cite{MW}.
However, this result 
contradicts the standard view, based on the work of Altshuler,
Aronov and Khmelnitskii\cite{AAK82} (AAK), that
%\begin{eqnarray}
%  \label{eq:AAKmainclaim}
$ \tauphi (T \to 0) = \infty$,
%\end{eqnarray}
and hence elicited a considerable and ongoing controversy
\cite{controversy}, with pertinent critique coming, 
in particular, from Ref.~\cite{AAG98,AAV01}.

The fact that  GZ's final results for $\tauphi (T)$ are controversial,
however, does not imply that their influence functional approach, as
such, is fundamentally flawed. To the contrary, having repeated their
calculations in detail, we have come to the conclusion that their
strategy is sound in principle and that  an influence functional  of
the form $e^{-(i \bar S_R + \bar S_I)}$ which they found can indeed be
derived without making non-standard approximations.  In fact,  it can
be shown, and this is our main result, that \emph{the standard Keldysh
  diagrammatic expressions for the self energy of the Cooperon can be
  obtained from} $i \bar S_R + \bar S_I$.  However,   when
\emph{applying} this influence functional to the problem of
decoherence, GZ make a semiclassically-motivated approximation
according to which the the effects of $\bsr$ for decoherence can be
neglected. We recount a simple  back-of-the-envelope
argument\cite{Marquardt-priv}, due to F. Marquardt, to suggest that
neglecting $\bsr$ is permissible  only for $T \gtrsim\hbar / \tauel$,
where $\tauel$ is the elastic mean free time. We shall also show that
in diagrammatic language,  neglecting $\bsr$  corresponds to
neglecting several important diagrams contributing to the Cooperon
self energy, as first pointed out in Ref.~\cite{AAV01}.

 Equation numbers from GZ's papers \cite{GZ1,GZ2,GZ3,GZS},
 will be prefaced, when
cited below, by I, II, III or IV, respectively.

%Due to lack of space, %and for the sake of readability of this paper,
%we shall not attempt to present here our rederivation of GZ's
%influence functional (detailed notes in electronic form can be made
%available upon request).  Instead, we  merely write down  the key
%results for the influence functional  (delegating technical
%definitions to an appendix), give a back-of-the-envelope
%argument\cite{Marquardt-priv}  for why GZ's approximate evaluation of
%its effects requires $T \gg \hbar / \tauel$, and state how the
%standard Keldysh diagrams for the  Cooperon self-energy can be recovered from
%$i \bar S_R + \bar S_I$.

\section{The model}

We consider a disordered system of interacting fermions,
with Hamiltonian $\hat H = \hat H_0 + \hat H_\irm$, where
\begin{align}
  \hat H_0 & =  \int d x \,
\hat \psi^\dag (x)  h_0 (x) \hat \psi (x) \; ,
%\\
% h_0 (x) &= \frac{- \hbar^2 }{2 m} \bnabla_{\bmr}^2 +
%V^\pdag_\imp (\bmr) - \mu \; ,
\\\nonumber %  \label{eq:defHint} \nonumber
\hat H_\irm  & =  {e^2 \over 2} \!\!
 \int \!\!   d x_1 \, d x_2 \,
: \! \hat \psi^\dag (x_1) \hat \psi (x_1) \! :
\tilde V^\inter_{12}
: \! \hat \psi^\dag (x_2) \hat \psi (x_2) \! :
\qquad \phantom{.}
%\\
%\hat n^\pdag_{\ij } & = 
%\hat \psi^\dag (x_j) \hat \psi (x_i) \; ,
%\\ \label{eq:defrho0}
% \langle \hat O \rangle_0  & = \Tr \{ \hat O \, \hat \rho_0 \}  ,
%\quad  \hat \rho_0 =  e^{- \beta \hat H_0} /
%   \{ \Tr e^{- \beta \hat H_0} \}  . \qquad \phantom{.}
\end{align}
Here $\int dx = \sum_\sigma \int d \bmr$,  and 
$\hat \psi
(x) \equiv \hat \psi (\bmr, \sigma) $ is the 
electron field operator for creating a spin-$\sigma$ electron at position
$\bmr$, with the following expansion in terms of the exact
eigenfunctions $\psi_\lambda(x)$ of 
$h_0 (x) = \frac{- \hbar^2 }{2 m} \bnabla_{\bmr}^2 +
V^\pdag_\imp (\bmr) - \mu$:
\begin{align}
\label{eq:definepsifields}
  \hat \psi (x)  =  \sum_\lambda \psi_\lambda (x) \hat c_{\lambda} ,
  \quad  \mbox{[} h_0 (x) - \xi_\lambda \mbox{]} 
\psi_\lambda (x) = 0. 
\end{align}
The interaction potential $\tilde V^\inter_{12} = \tilde V^\inter
(|\bmr_1 - \bmr_2|)$
 acts between the normal-ordered densities  at
$\bmr_1$ and $\bmr_2$.

\section{Influence functional for interacting electrons}

\begin{figure}[t]
\begin{center}
\includegraphics[scale=0.7]{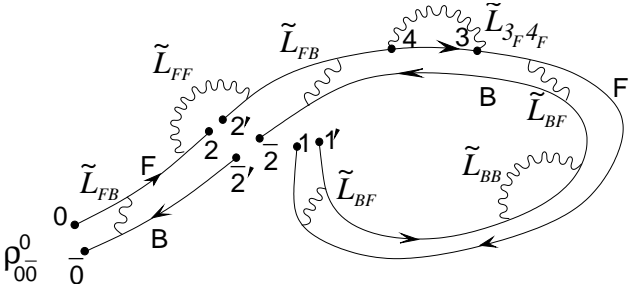}
\caption{A pair of paths contributing to the weak localization
  correction, illustrating 
  \Eq{eq:intermediateJrho}.  The interaction lines, generated by
  expanding the influence functional $e^{-(i \bsr + \bsi)}$
as in \Eq{eq:expandUU},  are
  labelled according to \Eq{eq:Ltildes}.
\label{fig:timereversedpaths}}
\end{center}
\end{figure}

GZ proposed a strategy (whose steps and approximations are
recapitulated in the Appendix),   that allows the DC conductivity,
$\sigma_\DC$, to  be expressed [see \Eq{eq:intermediateJrho}] in terms
of path integrals of the following general form [cf. (II.53), (IV.31)]:
\begin{align}
  \label{eq:J1221}
& \tilde {\cal C}_{12', {\bar 2}1'}
 =  \int_{2'_F, \bar 2_B}^{1_F, 1'_B}  {\cal D} (\bmR \bmP) 
 e^{-[i \bar S_R + \bar S_I](t_1,t_2)/\hbar} \, . \; 
\end{align}
%(The actual expression needed for $\sigma_\DC$, 
%\Eq{eq:intermediateJrho}, is more complicated, since it involves
%thermal averaging over initial coordinates
% at an initial time $t_0 \to
%- \infty$, but considering $ \tilde {\cal C}_{12', {\bar 2}1'}$ suffices to
%illustrate the essential aspects of the problem.) 
The symbol $\int {\cal D}(\bmR \bmP)$ is  a shorthand for the
following coordinate-momementa double path integral,\label{asymmetric}
\begin{align} \nonumber
& \int_{2'_F, \bar 2_B}^{1_F, 1'_B}  {\cal D} (\bmR \bmP) 
\; = \; \int_{\bmR^F (t_2)  = \bmr_{2'}}^{\bmR^F (t_1) =
\mib{r}_1}  {\cal D} \bmR^F (t_3) \int {\cal D} \bmP^F (t_3)
\\
& \times \! \int_{\bmR^B (t_{2})  = \bmr_{\bar 2}}^{\bmR^B (t_{1}) =
\mib{r}_{1'}} \!\! {\cal D} \bmR^B(t_3) \int {\cal D} \bmP^B (t_3)
e^{i [\bar S_0^F - \bar S_0^B] (t_1,t_2)/ \hbar} ,
\qquad \phantom{.} \label{eq:defineIntDRP}
\end{align}
which, when taken by itself, gives the  amplitude for a free electron
to propagate from $\bmr_{2'}$ at time $t_2$ to $\bmr_{1}$ at $t_1$,
times the amplitude for a free electron to propagate from $\bmr_{1'}$
at time $t_1$ to $\bmr_{2}$ at $t_2$  [corresponding to the loop parts
of the paths in Fig.~\ref{fig:timereversedpaths}], in the absence  of
interactions with other electrons.    We shall call these the forward
and backward paths, respectively, and label them by an index $a =
F,B$.   The corresponding free actions $\bar S_0^a = \bar S_0^{F/B}$
are given in \Eq{eq:defineS0-electrons}.  The weak localization
correction to the conductivity, $\sigma^\textrm{WL}_\DC$,  arises from
contributions  to $\sigma_\DC$ for which the coordinates $\bmr_1$,
$\bmr_1'$, $\bmr_2$ and $\bmr_2'$  all lie close together.  We
henceforth consider only this case. Then  $\tilde {\cal C}_{12', {\bar
    2}1'}$  is just the Cooperon propagator,  dominated by
contributions from those classical paths for which path $B$ is the
time-reversed version of path $F$.
The effect of the other electrons on this propagation is encoded in
the influence functional $e^{- (i \bar S_R + \bar S_I)}$ occuring in
\Eq{eq:J1221}. The effective action  $i \bar S_R + \bar S_I$ turns out
to have the form
%\begin{subequations}
\begin{align}
  \label{eq:SIR-LIR-aa}
   \bar  S_{R/I}  (t_1,t_2)  =  & \sum_{a,a'= F,B} 
\int_{t_2}^{t_1} d t_{3}  \int_{t_2}^{t_{3}} d t_{4}
 \, \bar L^{R/I}_{3_a 4_{a'}}  \; , 
\end{align}
where the $\bar L^{R}_{3_a 4_{a'}}$ are functions of the coordinates
and momenta $\bmR^a (t_3), \bmP^a (t_3)$ and $\bmR^{a'} (t_4),
\bmP^{a'} (t_4)$ that occur in the path integral [cf. (II.54,55)]:
\begin{subequations}
\label{eq:LRI}
\begin{align}
\nonumber
\bar L^{R}_{3_a 4_{a'}}    = &
 s_{a'} \tilde R \bigl(t_{3} - t_{4}, \bmR^a(t_{3}) - \bmR^{a'}  
(t_4)\bigr)  \qquad \phantom{.}\\
& \times 
\toh \bigl[ 1 - 2 \bar 
\rho_0^a ( \bmR^{a'} (t_{4}), \bmP^{a'} (t_{4}) \bigr]
\; ,     \label{eq:LRwaa}
\\
\bar L^{I}_{3_a 4_{a'}}    = &  
 s_a s_{a'}  
\tilde I \bigl( t_{3} - t_{4}, \bmR^a(t_{3}) - \bmR^{a'}  
(t_{4}) \bigr) \; , 
   \label{eq:Iwaa}
\end{align}
\end{subequations}
Here  $s_a$ stands for $s_{F/B} = \pm 1$, $\bar \rho^a_0(\bmR, \bmP)$
is the single-particle density matrix in a
mixed position-momentum represenation [cf. \Eq{eq:define-bar-rho}],
while $\tilde R(t, \bmR)$ and $\tilde I (t, \bmR)$ 
are real functions [given by \Eq{eq:defineRI}] that are,
respectively, proportional to retarded and Keldysh parts of
the interaction propatators ($\tilde R = {\cal L}^R$,
$\tilde I = i \toh {\cal L}^K$).

Via the influence functional,  \Eqs{eq:J1221} to (\ref{eq:LRI})
concisely incorporate the  effects of interactions  into the path
integral approach.  $\bar S_I$ describes  the \emph{classical} part of
the effective environment, and corresponds to the contribution
calculated by AAK\cite{AAK82}.  With  $\bar S_R$, GZ succeeded to
additionally also include the quantum part of the environment, and in
particular, via the occurence of the density matrix $\bar
\rho^a_0(\bmR, \bmP)$ in \Eq{eq:LRwaa},  to properly account for the
Pauli principle. Note, though, that \Eqs{eq:J1221} to (\ref{eq:LRI})
all refer to a given impurity configuration; impurity averaging still
has to be performed, and in the path integral formalism it is by no
means easy to do this properly.

\section{GZ's strategy for determining $\tauphi$}

To calculate the decoherence time $\tauphi$, 
GZ argue as follows: the effective action in \Eq{eq:J1221} in general
causes the Cooperon to decay with increasing time,  say as  $\tilde
{\cal C}_{12', {\bar 2}1'} \sim \tilde {\cal C}^0_{12', {\bar 2}1'}
e^{- f_d (t_1 - t_2)}$ [cf. (II.66), (IV.10)],   where $f_d (t)$
is an increasing function of time;  $\tau_\varphi$ is the time scale
characterizing this decay, set by $f_d (\tauphi) \simeq 1$.  To obtain
the function $f_d $,  GZ evoke  a  standard semiclassical argument:
since the path integral is dominated  by the saddle point paths of the
free action $\bar S^a_0$, \ie\ by the set of \emph{classical},
time-reversed diffusive paths, they take $f_d$ to  be (i) the disorder
average\cite{disorderaveraging} $\langle \; \rangle_\dis$  of the sum
over all classical paths $\langle \; \rangle_\classical$ of the
effective action evaluated along such a path $(i \src + \sic)$, but
(ii) without including any non-classical paths:
%be the effective
%action  evaluated along such a classical
%path ($i \src + \sic$) and then averaged
%over all such paths (denoted by $\langle \; \rangle_\classical$),
%but (i) without considering any non-classical paths, 
%and then (ii) averaged over disorder:
\begin{align}
\label{eq:fd}
 f_d (t_1 - t_2) \equiv  
\Bigl\langle \bigl\langle (i \src + \sic)(t_1, t_2)
\bigr\rangle_\classical \Bigr\rangle_\dis \; 
\end{align}
[cf. (III.22), (IV.11)].  Moreover, in the spirit of
semiclassical approximations, they (iii) take the limit\cite{hbarto0}
$\hbar \to 0$ in the single-particle density matrix occuring in
\Eq{eq:LRwaa} for $\bar S_R$, \ie, they replace $\bar \rho^a_0 (\bmR,
\bmP)$ of \Eq{eq:define-bar-rho} by the ``occupation number'' $n_0
\bigl(h_0 (\bmR, \bmP)\bigr)$, where  $n_0 (\xi) = 1/[e^{\xi/T} + 1]$
is  the Fermi function [cf.\ (II.43), (II.68)].

\emph{Within the approximations (i), (ii) and (iii),  GZ find that
  $\src \simeq  0$ for any given pair of classical, time-reversed
  paths, and hence conclude  that  $\bsr$ is ``irrelevant'' for
  decoherence, which is thus determined  by $\bsi$ alone}  [cf.\
discussion before (III.22), or after (IV.31).] They thus
calculate $f_d$ purely from $\sic$ [see, \eg, (IV.12)], and from
$f_d (\tauphi) \simeq 1$ find,  in 1 dimension for example,
[cf. (II.76,77)]
\begin{align}\label{eq:tauphi-general}
  \frac{1}{\tau_\varphi} &= \frac{ e^2 \sqrt{2D}}{\sigma_\DC}
\int_{1 / \tauphi}^{\omax} \frac{d \omega}{2 \pi}
\frac{ \coth (\omega/2T)}{\sqrt{\omega}}
\\
\label{eq:tauphi-0}
& =  \frac{ e^2 \sqrt{2D} }{\pi \sigma_\DC}
(2 T \sqrt{\tauphi} + \sqrt{\omax}\, )
\; . 
\end{align}
Note that the frequency integral has an ultraviolet divergence  at large
$\omega$, and hence has to be cut off by hand.\cite{complete} GZ cut
it off at $\omega^{\GZ}_{\max} = 1/\tauel$, the inverse elastic mean
free scattering time, arguing\cite{complete} that at higher $\omega$
(smaller times) the ``approximation of electron diffusion becomes
incorrect'' [cf. paragraph before (II.76)]. This leads to a finite
decoherence time at zero temperature in \Eq{eq:tauphi-0},
$1/ \tauphi (T \to 0) \neq 0$.  In contrast, according to the
philosophy of AAK \cite{AAK82}, one should take $
\omega^\textrm{AAK}_{\max} = T/\hbar$, since frequencies larger than
$T/\hbar$ would correspond to \emph{virtual} excitations of the
environment, which are believed not to contribute to decoherence.
This would yield the standard result  $1/\tauphi (T
\to 0)= 0$.  Thus, \emph{the controversy centers on the
  question whether the higher frequency modes do contribute to
low temperature   decoherence or not}.

\section{On the importance of $\bar S_R$ for decoherence}
\label{sec:SRimportance}

Although we agree with GZ's influence functional
(\ref{eq:J1221}) to (\ref{eq:LRI}), \emph{we disagree with GZ's central
conclusion that $\bsr$ is irrelevant for decoherence.}   Firstly, 
 influence functionals have the general feature  that  deducing
decoherence properties from $\bsi$ alone is reliable only for highly
excited states: as emphasized in a very insightful recent paper by
F. Marquardt \cite{Marquardt02}, neglecting $\bsr$ would amount to
neglecting dissipative effects by which the quantum system can give
back to the environment some of the  energy which it picks up from the
classical part of the environment; this neglect of dissipation
would thus cause the system to heat up beyond what is allowed by
detailed balance  (an effect which can be ignored for decoherence only
if the system is already highly excited).

Secondly, and more specifically: even if  one puts aside  all
reservations\cite{hbarto0,disorderaveraging} about approximations (i)
and (iii) [although these reservations \emph{are} serious], and
accepts GZ's result that $\src \simeq 0$ for classical paths, a
question remains about approximation (ii): what about fluctuations,
\ie\ contributions from paths in the path integral that deviate
slightly from the classical paths''?  If the ``leading'' contribution
vanishes, the ''subleading'' one becomes of interest!  Indeed, it is
natural to expect  that such close-to-classical paths can produce a
significant contribution missed by GZ's  semiclassics, say $\dsr \neq 0$,
 due to the following intrinsic problem (IP): 
 \emph{$\bsr$
  depends on the density matrix $\bar \rho_0(\bmR, \bmP)$, which at low
  termperatures is a very sharp function of $\bmR$ and $\bmP$, and
  hence very sensitive to fluctuations of  (or approximations
  inside~\cite{hbarto0}) its arguments}. 
This intrinsic problem sets  the present
problem apart from other, exactly solvable models
such as the Caldeira-Leggett model, and renders meaningless any
attempts\cite{GZ3} to justify neglecting $\bsr$ here by comparisons to such
other models.

Let us estimate\cite{Marquardt-priv} the importance of these
fluctuations, by estimating  $\dsr$ for a segment of a diffusive path
in which an electron travels ballistically with velocity $\vF$ (and
energy $\simeq \mu$) for a time $\tauel$ between two scattering events
at two impurities. Now, the close-to-classical paths that still
contribute significantly to the path integral are those for  which
$(\bar S_0 -  \bar S_0^\classical)  \lesssim \hbar$.  Considering only
the first term in \Eq{eq:defineS0-electrons} for $\bar S_0$, we thus use
\begin{align}
\label{eq:estimatedeltap}  
\hbar \simeq  (\bar S_0 -  \bar S_0^\classical) \simeq \int^\tauel_0 d
t \,  \delta \bmP \cdot \dot \bmR   \simeq \tauel \, \delta
P  \, \vF,
 \end{align}
 or $\delta P \simeq \hbar / \tauel \vF$ (which is just the
 uncertainty principle), to estimate the typical magnitude of momentum
 fluctuations around $P_\textrm{F}$.  It follows that
\begin{align}
  \label{eq:deltan}
\delta n_0 \bigl( h_0 (\bmR, \bmP) \bigr) \simeq
\frac{\delta  h_0 (\bmR, \bmP)}{T} =
\frac{P_\textrm{F} \delta P}{m T} = \frac{\hbar}{\tauel T}.
\end{align}
Now, \emph{GZ neglect  $\dsr$ relative to $\sic$; since this
  requires $\delta n_0 \lesssim 1$, it  
can be reliable only for $T \gtrsim \hbar / \tauel$.} 
 For smaller $T$,
  the contributions of $\dsr$ should  become
  important.\cite{consistencychecks}

Having realized that $\dsr$ is important  at low
temperatures, the  natural next question is: can $\dsr$ cancel the
ultraviolet divergence arising from $\sic$? We believe it does, but
showing this will require  a more accurate calculation  than GZ's,
that does not resort to semiclassical arguments and avoids
approximations (i) to (iii).

\section{Obtaining diagrams from influence functional}
\label{sec:obtainingaa}

\begin{figure}[t]
\begin{center}
\includegraphics[scale=0.35]{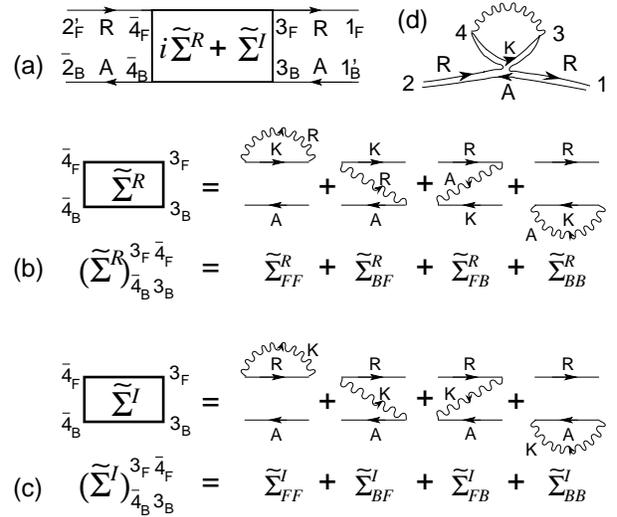}
\caption{\label{fig:selfenergy}
First order contributions to the irreducible self energy
of the Cooperon, illustrating (a) \Eq{eq:selfenergyN=1},
and (b,c) \Eqs{eq:selfenergies-explicit}. (d): 
typical path contributing to $\tilde \Sigma^R_{FF}$;
GZ neglect $\tilde \Sigma^R$ and hence such paths,
as first pointed out in Ref.~\cite{AAG98}.}
\end{center}
\end{figure}

As a first step in that direction, we have  explored the connection
between GZ's influence functional  and  standard diagrammatic
perturbation theory. The connection turns out to be remarkably simple:
upon performing the momentum integrals in $\int {\cal D} (\bmR \bmP)$
and expanding  the resulting influence functional (some details are
given in App.~A.2),  one generates a Dyson-like
equation for the Cooperon [\Eq{eq:expandUU}], with a self energy whose
lowest order\cite{othercorrections} irreducible diagrams  [given by
\Eq{eq:selfenergies-explicit}] are depicted diagrammatically in
Fig.~\ref{fig:selfenergy}.  Remarkably, \emph{the resulting diagrams
  coincide precisely with those obtained by standard Keldysh
  diagrammatic perturbation theory,} as depicted, \eg, in Fig.~2 of
Ref.~\cite{AAV01} (There, impurity lines needed for impurity averaging
are also depicted;  in our Fig.~\ref{fig:selfenergy}, they are
suppressed).  This fact, which is our main new result, is a strong
indication that the  expressions  of \Eqs{eq:J1221} to (\ref{eq:LRI})
for the influence functional are sound, and the approximations made
during its derivation reasonable [steps (5) and (6) in App.\ A.1, and
Ref.~\cite{Hartree}].

Moreover, this fact also allows us to shed new light on the roles of
$\bsr$ and $\bsi$ in the effective action. $\bsi$ gives rise  to the
terms $\tilde \Sigma_I$, which contain a factor  $\coth (\hbar \omega
/ 2T)$ [arising from the factor $\bar I (\bmk, \omega)$, cf.\
\Eqs{eq:defineI-wk} and (\ref{eq:Ltildes})], and $\bsr$ gives rise  to
the terms $\tilde \Sigma_R$, which contain a factor
$\tanh[(\varepsilon_\lambda - \omega) / 2T]$  [arising from the factor
$(\tilde \delta - 2 \tilde \rho^0)$ in \Eq{eq:Ltildes}].  Now, the sum
of all the self-energy diagrams for $i \tilde \Sigma_R + \tilde
\Sigma_R$ in Fig.~\ref{fig:selfenergy}  has been evaluated in
Ref.\cite{AAV01}, in the energy-momentum representation and for those
choices of energy and momentum arguments that are relevant for
determining the Cooperon lifetime; the result [Eqs.~(3,4) of
Ref.~\cite{AAV01}] was found to be \emph{not} ultraviolet divergent,
since the $\coth$ and $\tanh$ functions always occur in the combination
\begin{align}
\label{eq:tanh-cothcanel}
  \int \frac{d \omega}{2 \pi} \left\{ 
\tanh[(\varepsilon_\lambda - \omega) / 2T]
+  \coth (\omega / 2T) \right\}  \dots\; ,
\end{align}
so that the frequency integral \emph{is} cut off at $\omax \simeq T/
\hbar$, as anticipated by AAK\cite{AAK82}.  We expect this
cancellation of UV divergencies from $i \bsr$ and $ \bsi$  to occur
not only in first order perturbation theory, but in every
order,\cite{SRSIcannotcancel} since the structure,  $e^{-(i \bsr +
  \bsi)}$, of the influence functional is such that  the self energy
contribution $i\tilde \Sigma_R$ and $ \tilde \Sigma_I$ \emph{always}
occur in the combination $i\tilde \Sigma_R + \tilde
\Sigma_I$.\cite{othercorrections}

Conversely, we conclude that GZ's neglect of $\dsr$ corresponds to
neglecting the contribution of $\bsr$ to the Cooperon self energy,
\ie, \emph{GZ neglect all the  diagrams of
  Fig~\ref{fig:selfenergy}(b),} as first pointed out in
Ref.~\cite{AAV01}.  As argued above, we believe that this is allowed
only if $T \gtrsim \hbar / \tauel$.

\section{Conclusion}

Our analysis can be summarized as follows:   GZ's influence functional
stratey is sound in principle;  when implemented with sufficient
care, it properly incorporates, via $\bsr$, the Pauli principle. However,
GZ neglect the latter by neglecting $\dsr$
%the beauty of their
%influence functional is that it properly incorporates the Pauli
%principle, via $\bsr$. However, the spoils of this achievement are
%lost as soon as $\dsr$, and hence the Pauli principle, are neglected
during the semiclassical calculation of  $\tauphi$. This can work only
for large temperatures.

A complete, first-principle evaluation of $\tauphi$ would be obtained
if one sums up the Dyson-like equation (\ref{eq:expandUU}) for the
Cooperon  in the presence of disorder, either diagrammatically or by
using path integral techniques, but \emph{without} neglecting
$\dsr$. To the best of our knowledge,  this program has not yet been
fully carried out to the end.  Thus, at least in the eyes of the
present author, the fate of $\tauphi$ as $T\to 0$ for the present
model has not yet been worked out in complete, conclusive detail.  At
present we see no reason to believe, though, that the result will
disagree with the conclusions of AAK.

\section*{Acknowledgements}
I thank  D. Golubev and A. Zaikin, and, in equal measure, I. Aleiner,
B. Altshuler, M. Vavilov, I. Gornyi and F. Marquardt, for numerous
patient and constructive discussions,  which taught me many details
and subtleties of the influence functional and diagrammatic
approaches,  and  without which I would never have been able to reach
the conclusions presented above.   Moreover, I  thank V. Ambegaokar
for raising and sustaining my interest in this problem, and
acknowledge illuminating discussions with J. Imry, J. Kroha,
S. Mirlin, H. Monien, A. Rosch, I.  Smolyarenko, G. Sch\"on,
P. W\"olfle and A.  Zawadowski.

\emph{Note added in proof:---} Very recently, 
GZ defended their theory yet again\cite{GZ5} in cond-mat/020814.
%and were critized yet again, in cond-mat/0208264.

\appendix
\section{Appendix}

This appendix summarizes, without derivations,  useful technical
results that are alluded to in the main text.

\subsection{Kubo formula in terms of path integrals}

The Kubo formula for the DC conductivity of a $d$-dimensional
conductor can be expressed as
%in either of  the forms
%\begin{subequations}
%\label{eq:sigmaDC-final}
\begin{align}
\nonumber
\sigma_\textrm{DC} & =  -     \lim_{\omega_0 \to 0}
{e \over d } \sum_{\sigma_1}
%\sum_{\alpha}
\int \! dx_2 \,
    \mib{j}_{11'} \cdot  \bmr_2
    \tilde {   G}^R_{11',22} (\omega_0)\Big|_{x_1 = x_{1'}} , \,
\end{align}
%\\
%\label{eq:sigmaAAG}
%\sigma_\textrm{DC} & =  
%    \lim_{\omega_0 \to 0} {1 \over \omega_0} \, \sum_{\sigma_1}
%\left[ \frac{i \, e^2}{m  }\langle \hat \psi^\dag (t_1, x_1) \hat \psi (t_1,
%    x_1)  \rangle  \right. \\
% &  \qquad \left. \nonumber
%+ {1 \over d } 
%\int \!  dx_2 \, \mib{j}_{11'} \cdot \mib{j}_{\,22'}
%\tilde {   G}^R_{11',22'}
%    (\omega_0)\Big|_{\stackrel{\scriptscriptstyle 
%{x_1 = x_{1'}}}{\scriptscriptstyle {x_2 = x_{2'}}}} , \,  
%\right] \;, \qquad \phantom{.}
%\end{align}
%\end{subequations}
%depending on  whether one chooses, respectively, to represent a 
%uniform applied
%electric field $\mib{E}(\omega_0)$ using a scalar potential $V_\ext
%(\omega_0,\bmr_2) = - \bmr_2 \cdot \mib{E}(\omega_0)$  (as GZ
%do\cite{GZ2}), or  a vector potential $\mib{A}_\ext (\omega_0) =
%\mib{E} (\omega_0)/i \omega_0$ (as AAG do\cite{AAG97}).
%  (as GZ
%do\cite{GZ2}), or  a vector potential $\mib{A}_\ext (\omega_0) =
%\mib{E} (\omega_0)/i \omega_0$ (as AAG do\cite{AAG97}).
\begin{align}
\label{eq:defineGomega0}
\tilde {   G}^R_{11',22'} (\omega_0) &= 
 \! \int_{- \infty}^\infty
\!\!  dt_{12} e^{i \omega_0 t_{12}}
 \theta (t_{12}) \, \tilde {   C}_{[11',22']} \; ,
\end{align}
\begin{align}
\nonumber %\label{eq:define-C-firsttime-a}
\tilde {\cal C}_{[11',22']}  \equiv  {1 \over \hbar} 
\langle [\hat \psi^\dag (t_1,x_{1'}) \hat \psi (t_1,x_{1}), \hat  \psi^\dag
(t_2,x_{2'}) \hat \psi (t_2,x_{2}) ] \rangle_H , \qquad \phantom{.}
\end{align}
where $ \mib{j}_{11'}  \equiv {- i e \hbar \over 2 m}
( \bnabla_1 - \bnabla_{1'})$ and a  uniform applied
electric field $\mib{E}(\omega_0)$ was
represented using a scalar potential, $V_\ext
(\omega_0,\bmr_2) = - \bmr_2 \cdot \mib{E}(\omega_0)$.
%Both the time dependence and the statistical weighting in the
%generalized density-density commutator $\tilde {   C}_{[11',22']} $
%are governed by the full Hamiltonian $\hat H$.  Note that the indices
%1 and $1'$ refer to the same time $t_1$, and likewise for 2, $2'$ and
%$t_2$. 
A path integral representation for $\tilde {   C}_{[11',22']} $
can be derived using  the following strategy, adapted\cite{deviations}
from GZ's Ref.~\cite{GZ2}: (1) introduce a source term, in which an
artificial source field $\tilde v_{2'2}$  couples to $ \hat \psi^\dag
(t_2,x_{2'}) \hat \psi (t_2,x_{2})$, and write $\tilde {
  C}_{[11',22']} $ as the linear response of the single-particle
density matrix  $\tilde {\mib{\rho}}_{11'} =  \langle \hat \psi^\dag
(t_1,x_{1'}) \hat \psi (t_1,x_{1}) \rangle_H$ to the source field
$\tilde v_{22'}$.  (2) Decouple the interaction using a
Hubbard-Stratonovitch transformation, thereby introducing a functional
integral over real scalar fields $V_{F/B}$, the so-called
``interaction fields''; these then  constitute a dynamic, dissipative
environment with which the electrons interact. (3) Derive an equation
of motion for $\tilde \rho_{11'}^V$, the single-particle density
matrix for a given, fixed configuration of the fields $V_{F/B}$, and
linearize it in $\tilde v_{2'2}$, to obtain an equation of motion for
the linear response $\delta \tilde \rho_{11'}^V (t)$ to the source
field.  (4) Formally integrate this  equation of motion  by
introducing a path integral $\int {\cal D} (\bmR \bmP)$ over the
coordinates and momenta of the single degree of freedom associated
with the single-particle density matrix $\delta \tilde
\rho^V_{11'}$. (5) Use the RPA-approximation to bring the effective
action $S_V$ that governs the dynamics of the fields $V_{F/B}$ into a
quadratic form. (6) Neglect the effect of the interaction on the
single-particle density matrix wherever it still occurs, i.e.\
replace $\tilde \rho^V_{\ij}$ by the free single-particle density
matrix
\begin{align}
\label{eq:exact-rho0-a}
\tilde  \rho^0_\ij   = 
 \langle \hat \psi^\dag (x_j) \hat \psi (x_i) \rangle_0  
=  \sum_\lambda \psi^\ast_\lambda (x_j) \psi_\lambda (x_i) \, n_0
 (\xi_\lambda)
\; ,
\end{align}
where $\langle \hat O\rangle_0  = \Tr e[^{- \beta \hat H_0} \hat O ]
/ \Tr [e^{- \beta \hat   H_0}]$. 
(7) Perform the functional integral (which steps (5) and
(6) have rendered Gaussian) over the fields $V_{F/B}$; the environment
is thereby integrated out, and its effects on the dynamics of the
single particle are encoded in an influence functional of the form
$e^{-(i \bar S_R + \bar S_I)}$. 
The final result of this strategy is that 
 $\tilde {\cal C}_{[11',22']}$ can be written as\cite{Hartree}
[cf. (II.49)]
\begin{align}
\nonumber
% \label{eq:C1122afterderivatives}
%&    \tilde {\cal C}_{[11',22']}
% =    \\ & 
& \lim_{t_0 \to - \infty} \int \! d x_{\bar 2} \,
  \nonumber
 \{ 
\tilde J_{12', 2 \bar 2, \bar 2 1'}(t_1,  t_2; t_0)  
 -   \tilde J_{1 \bar 2, \bar 2 2', 21'}(t_1, t_2; t_0) \},
\\
%\end{align}
%\begin{align}
  \label{eq:intermediateJrho}
& \tilde J_{12',2 \bar 2', {\bar 2}1'} (t_1,t_2; t_0)
\rangle_V^\pdag = \int_{2'_F, \bar 2_B}^{1_F, 1'_B}  {\cal D} (\bmR \bmP) 
\\ & \nonumber \quad
\times \! 
\int_{0_F, \bar 0_B}^{2_F, \bar 2'_B}  {\cal D} (\bmR \bmP) 
 \frac{ 1}{\hbar}
\int d x_0  d x_{\bar 0} \, \tilde \rho^0_{0 \bar 0} \, 
 e^{-[i \bar S_R + \bar S_I](t_1,t_0)/\hbar} \, .
\end{align}
where  $t_{\bar 2} = t_2$, i.e.\  the indices 2 and $\bar 2$ refer to
the same time  $t_2$, but  the integration variable $x_{\bar 2}$  is
independent of $x_2$.
%, and $\tilde J$ is given by the following  path integral:
 The symbol $\int {\cal D} (\bmR \bmP)$ stands
for the double path integral defined in \Eq{eq:defineIntDRP}. The
complex weighting functional $e^{i (\bar S_0^F - \bar S_0^B)}$
occuring  therein involves  the  action for a single, free electron,
in the mixed coordinate-momentum representation,
\begin{align}
  \label{eq:defineS0-electrons}
&  \bar S_0^a (t_i,t_j)[\bmR^a (t_3), \bmP^a(t_3)]
\\ & \quad \nonumber
 = \int_{t_j}^{t_i} d t_3
\left[ \bmP^a (t_3) \cdot \partial_{t_3} \bmR^a (t_3) - 
\bar h_0 \bigl(\bmR^a (t_3), \bmP^a (t_3) \bigr)
\right]  ,
\\  \label{eq:define-bar-h0}
& \bar h_0 (\bmR^a, \bmP^a) = { \bmP^{a2} \over 2m} +
V_\text{imp}(\bmR^a) - \mu \;  .
\end{align}

The first line of \Eq{eq:intermediateJrho}  corresponds to
\Eq{eq:J1221}, and its significance as describing Cooperon propagation
along two paths from  $\bmr_{2'}$ at time $t_2$ to $\bmr_{1}$ at
$t_1$, and $\bmr_{1'}$ at time $t_1$ to $\bmr_{2}$ at $t_2$, is
explained after \Eq{eq:defineIntDRP}. The integrals in the second line
of \Eq{eq:intermediateJrho}  ``initialize'' this
Cooperon\cite{t_0} at an initial time $t_0 \to - \infty$, by including
forward and backward propagation [see the left (non-loop) part of
Fig.~\ref{fig:timereversedpaths}] between the times $t_0$ and $t_2$,
from and to some initial positions $\bmr_0$ and $\bmr_{\bar 0}$, which
are  weighted by the initial, free single-particle density matrix
$\tilde \rho^0_{0 \bar 0}$ of \Eq{eq:exact-rho0-a}.

The effective action   $i \bar S_R + \bar S_I$
in the influence functional $e^{- (i \bar S_R + \bar S_I)}$ in
\Eq{eq:intermediateJrho} is given by
\Eqs{eq:SIR-LIR-aa} and (\ref{eq:LRI}).  
$\bar  S_R$  depends via \Eq{eq:LRwaa} on the
single-particle density matrix  in a mixed coordinate-momentum
representation,\cite{asymmetric}
\begin{align}
\nonumber
\bar \rho^F_0 (\bmR^F, \bmP^F) & =  \int d \bmr^F \, 
 e^{- i   \bmp^F \cdot \delta \bmr^F} 
\,  \tilde \rho^0 (\bmR^F + \delta \bmr^F, \bmR^F ) \; ,
\\
 \label{eq:define-bar-rho}
\bar \rho^B_0 (\bmR^B, \bmP^B) & =  \int d \bmr^B \, 
 e^{- i   \bmp^B \cdot \delta \bmr^B} 
 \,  \tilde \rho^0 (\bmR^B , \bmR^B - \delta \bmr^B) \; 
\end{align}
where $\tilde \rho^0(\bmr_i, \bmr_j) = \tilde \rho^0_\ij$
is defined in \Eq{eq:exact-rho0-a}.

Furthermore, $i \bar S_R + \bar S_I$ depends on two purely real
functions,  $\tilde R(t,\bmr) $ and $\tilde I(t, \bmr)$,  that  are
defined  as follows via their Fourier transforms [cf.(II.56,57)]:
\begin{subequations}   \label{eq:defineRI}
\label{eq:definegreensfunctions}
 \begin{align}
\nonumber % \label{eq:defineR-tr}
 ( \tilde R / \tilde I ) (t, \bmR) \! & = 
\int \! { d \bmk  d \omega  \over (2 \pi )^{d+1} } 
 e^{-i \omega t + {i} \bmk \cdot \bmR}
 (\bar R/\bar I) (\omega, \bmk) \; , \qquad \phantom{.}
%\end{align}
\\ 
% \begin{align} 
\label{eq:defineR-wk}
 \bar R(\omega, \bmk) &=
{\bar V^\inter (\bmk) \over 1 -  \bar V^\inter (\bmk) 
\bar \chi ( \omega, \bmk)}  ,
\\ \label{eq:defineI-wk}
\bar I(\omega, \bmk)
&=  - \coth (\hbar \omega /2T ) \, \textrm{Im} \, \bar R(\omega, \bmk) \; .
\end{align}
\end{subequations}
Here $\bar V^\inter (\bmq)$ and $\bar \chi ( \omega, \bmk) $
are the Fourier transforms of the
interaction potential $\tilde V^\inter (|\bmr_1 - \bmr_2|)$
%[For example, for a screened interaction of range $1/\lambda$,  we have
%\begin{eqnarray}
%  \label{eq:screenedCoulomb}
%\tilde V^\inter_{12} = { 
%e^{- \lambda |\bmr_1 - \bmr_2| } \over |\bmr_1 -
% \bmr_2|} , \qquad
%\bar V^\inter (\bmq) = {4 \pi \over \lambda^2 + \bmq^2} \; . 
%\end{eqnarray}
%GZ use the unscreened version of \Eq{eq:screenedCoulomb}, with
%$\lambda = 0$.] 
and  the charge susceptibility
$  \tilde \chi_{12}$, which can be written as 
\begin{align}
\label{eq:chietaGF}
  \tilde \chi_{12} & =  
- i  \, e^2  \hbar 
 \bigl( \G^R_{12} \G_{21}^K +
\G_{12}^K \G^A_{21} \bigr) \; .
\end{align}
$\tilde G^{R/A/K}_\ij$ are the retarded, advanced and Keldysh Green's
functions of the noninteracting, disordered system:
\begin{subequations}
\begin{align}
&  \G^{R/A}_{\ij}  = \mp {i\over \hbar} \theta(\pm t_\ij) \sum_\lambda
\psi^\ast_\lambda (x_j) \psi_\lambda (x_i)
e^{-i \xi_\lambda t_\ij/ \hbar } \; ,
\\ \nonumber 
& \G^K_{\ij}  = - { i \over \hbar} \!\! \sum_\lambda
\psi^\ast_\lambda (x_j) \psi_\lambda (x_i)
e^{-i \xi_\lambda t_\ij/ \hbar} [1 - 2 n_0 (\xi_\lambda )] 
\; . 
\end{align}
\end{subequations}

\subsection{Derivation of Cooperon self energy from
influence functional}

The path integrals $\int {\cal D} (\bmR \bmP)$ of
\Eq{eq:J1221} can be given a precise definition in terms of 
the standard time-slicing procedure for path integrals,
with one coordinate and one momentum integral
for every time slice.\cite{asymmetric} 
For each time slice (labeled by $n$, say),
the momentum integral $\int d \bmP_n$ 
can then easily be  performed, since it simply has the effect
of converting the expressions occuring in the action
for that time slice
from the mixed coordinate-momentum representation
to the coordinate-only representation. Thus, 
the free action  $\bar S^a_0 [\bmR, \bmP]$ is mapped to
\begin{align}
%\label{eq:standardfreeactionRonly}
\tilde S^a_0 (t,t') [\bmR^a, \dot \bmR^a]& =   \int_{t'}^{t}   dt_3
%\tilde L^a_0 \bigl(t_3, \bmR^a (t_3) \bigr) & \equiv &
\left[\toh m \dot \bmR^{a2} (t_3) - V_\imp \bigl( \bmR^a (t_3)
  \bigr)\right]  ,
\nonumber
\end{align} 
the density matrix at time slice $n$ is
converted\cite{asymmetric}  from $  \bar \rho^a_0(\bmR^a_n, \bmP^a_n)
$ to $\tilde \rho^0 (\bmr_n^a, \bmr_{n-1}^a)$, and $\tilde {\cal
  C}_{12', {\bar 2}1'}$ of \Eq{eq:J1221}  can be rewritten as
\begin{align}
\label{eq:JrhoRonly}   
& \tilde {\cal C}_{12', {\bar 2}1'} \; = \; 
 \int_{2'_F, \bar 2_B}^{1_F, 1'_B}   {\cal D}'\!  (\bmR)
\,  e^{- [i \tilde S_R + \tilde S_I](t_1,t_2)/ \hbar}  \; ,
\end{align}
where the integral $\int  {\cal D}' \! (\bmR )$
is used as a shorthand  for 
\begin{align}
\label{eq:coordonlyPI}
 \int_{2'_F, \bar 2_B}^{1_F, 1'_B}   {\cal D}' \!  (\bmR) & 
\; = \;  \int_{\bmR^F (t_2)  = \bmr_{2'}}^{\bmR^F (t_1) =
\mib{r}_1}   {\cal D}' \!  \bmR^F (t_3)
\\ \nonumber
\times &
 \int_{\bmR^B (t_2)  = \mib{r}_{\bar 2}}^{\bmR^B (t_1) =
\bmr_{1'}}  {\cal D}' \!  \bmR^B(t_3)
e^{i [\tilde  S_0^F - \tilde S_0^B]/ \hbar}  \; .
\qquad \phantom{.}
\end{align}
%The tildes on $\tilde S^a_0$, $\tilde S_{R/I}$ (and $\tilde L^{R/I}$
%below)
%denote that they are functionals of the paths $\bmR^a$ and
%velocities $\dot \bmR^a$  (but not of the
%momenta $\bmP^a$, which have been integrated out).
The effective action $i \tilde S_R + \tilde S_I$
in \Eq{eq:JrhoRonly} is found to 
have the following form\cite{xbarxintegrals}
[with  $\tilde \delta_{\bari i} = \delta_{\sigma_\bari
  \sigma_i} \delta (\bmr_\bari - \bmr_i)$  and
$(\tilde R/\tilde I)_{\bari_a \barj_{a'}} =  (\tilde R/\tilde I)
\bigl( t_i-t_j, \bmr^a_{\bari} (t_i)- \bmr^{a'}_\barj (t_j) \bigr) $]:
\begin{align}
\label{eq:Gaussian-Integral-Ronly}
\tilde S_{R/I} (t_1,t_2)   =  \sum_{aa'} \int_{t_2}^{t_1} d t_{3_a}
\int_{t_2}^{t_{3_a}} d t_{4_{a'}} \tilde L^{R/I}_{3_a 4_{a'}}  \; ,
\end{align}
\begin{subequations}
    \label{eq:Ltildes}
\begin{align}
  \label{eq:LtildesFF}
  \tilde L^{R/I}_{3_F 4_F}  & =  
\left \{ \begin{array}{l} 
\phantom{-} \toh \, \tilde \delta_{3_F \bar 3_F}
  (\tilde \delta - 2 \tilde \rho^0)_{4_F \bar 4_F} 
\, \tilde R_{\bar 3_F \bar   4_F}
\rule[-3mm]{0mm}{0mm}
\\
\phantom{- \toh} \,  \tilde \delta_{3_F \bar 3_F}  
\, \tilde \delta_{4_F \bar 4_F} 
\, \tilde I_{\bar 3_F \bar   4_F} \; , 
 \end{array} \right. \rule[-7mm]{0mm}{0mm}
\\
    \label{eq:LtildesBF}
\tilde L^{R/I}_{3_B 4_F}  & =  
\left \{ \begin{array}{l} 
 - \toh \, \tilde \delta_{\bar 3_B 3_B}
  (\tilde \delta - 2 \tilde \rho^0)_{4_F \bar 4_F} 
\, \tilde R_{\bar 3_B \bar   4_F}
\rule[-3mm]{0mm}{0mm}
\\
-  \phantom{ \toh\,}  \tilde \delta_{\bar 3_B  3_B}  
\, \tilde \delta_{4_F \bar 4_F} 
\, \tilde I_{\bar 3_B \bar   4_F} \;  , 
 \end{array} \right. \rule[-7mm]{0mm}{0mm}
\\
    \label{eq:LtildesFB}
\tilde L^{R/I}_{3_F 4_B}  & =  
\left \{ \begin{array}{l} 
\phantom{-} \toh \, \tilde \delta_{3_F \bar 3_F}
  (\tilde \delta - 2 \tilde \rho^0)_{\bar 4_B 4_B} 
\, \tilde R_{\bar 3_B \bar   4_B}
\rule[-3mm]{0mm}{0mm}
\\
- \phantom{ \toh\,}  \tilde \delta_{3_F \bar 3_F}  
\, \tilde \delta_{\bar 4_B 4_B} 
\, \tilde I_{\bar 3_F \bar   4_B} \; , 
 \end{array} \right. \rule[-7mm]{0mm}{0mm}
\\
    \label{eq:LtildesBB}
\tilde L^{R/I}_{3_B 4_B}  & =  
\left \{ \begin{array}{l} 
- \toh \, \tilde \delta_{\bar 3_B 3_B}
  (\tilde \delta - 2 \tilde \rho^0)_{\bar 4_B 4_B} 
\, \tilde R_{\bar 3_B \bar   4_B}
\rule[-3mm]{0mm}{0mm}
\\
\phantom{- \toh\, }  \tilde \delta_{\bar 3_B 3_B}  
\, \tilde \delta_{\bar 4_B 4_B} 
\, \tilde I_{\bar 3_B \bar   4_B} \; .
 \end{array} \right. \rule[-7mm]{0mm}{0mm}
\end{align}
\end{subequations}

Now expand  the effective action in powers of $\tilde S_{R/I}$,
\begin{align}  \label{eq:expandUU}
& \tilde {\cal C}_{12', {\bar 2}1'} 
 =  \sum_{N=0}^\infty {1 \over N!} 
\int_{2'_F, \bar 2_B}^{1_F, 1'_B}   {\cal D}' (\bmR)
\\  \nonumber
&  \times \left[ {- 1  \over \hbar}
\sum_{aa'}
\int_{t_0}^{t_1} \! \! d t_{3_a} \int_{t_0}^{t_{3_a}} \! \! d t_{4_{a'}} 
\bigl[ i \tilde L^{R}_{3_a 4_{a'}} + \tilde L^I_{3_a 4_{a'}} \bigr] 
 \right]^N \; , 
\end{align}
and analyze the structure of this expansion, which can be viewed as a
type of  Dyson equation for the Cooperon. The $N=0$ term  simply
yields free propagation between the specified  end points,
%\begin{align}
%  \label{eq:int0-RARA}
$\tilde {\cal C}_{12', {\bar 2}1'}^0 
= \hbar^2 \tilde G^{R , 1_F 2'_F}
 \tilde G^{A}_{\bar 2_B \bar 1'_B} \; .
$  %\end{align}
%where  for clarity, indices associated with the
%forward $(F)$ or backward $(B)$ paths are written as
%superscripts or subscripts, respectively.  
The $N=1$ term of \Eq{eq:expandUU} turns out  to have the form  of a
forward-backward propagator sandwidching  self-energy
insertions,\cite{xbarxintegrals} say $\tilde \Sigma^{R/I}_{aa'}  $,
that are depicted
     diagrammatically in      Fig.~\ref{fig:selfenergy}
and are given by:
\begin{align}
 & \nonumber 
\int_{2'_F, \bar 2_B}^{1_F, 1'_B}   {\cal D}' (\bmR) 
\tilde  L^{R/I}_{aa'} (t_3, t_4)
=  
\int  d x_{3_F} d x_{3_B} d x_{\bar 4_F} d x_{\bar 4_B}  
\quad \qquad \phantom{.}
\\
& \quad \label{eq:selfenergyN=1}
\hbar^4 \tilde G^{R, 1_F 3_F} \tilde G^{A}_{\bar 2_B \bar 4_B} 
 \left( \tilde \Sigma^{R/I}_{aa'} \right)^{3_F \bar
  4_F}_{\bar 4_B 3_B} 
\tilde G^{R, \bar 4_F 2'_F} \tilde G^{A}_{3_B 1'_B} \, .
%\\
%& 
%(\tilde U^F_B)^{\bari_F j_F}_{j_B \bari_B}  \equiv  
% \tilde U^{0F, \bari_F j_F} \tilde U^{0B}_{j_B \bari_B} \, , 
%\hbar^2 \tilde G^{R, \bari_F j_F} \tilde G^{A}_{j_B \bari_B} \, , 
\end{align}
%\begin{subequations}
  \begin{align} 
\nonumber %   \label{eq:SigmaFF}
\left( \tilde \Sigma^{R/I}_{FF} \right)^{3_F \bar   4_F}_{\bar 4_B 3_B} &= 
\phantom{-} 
%\left \{ \begin{array}{l} \!\!\! \toh \!\!\! \\ 
%\!\! 1 \!\! \end{array} \right\}
 \hbar^2 
(\tilde G^{K/R})^{3_F \bar 4_F} \tilde G^A_{\bar 4_B 3_B}
 ({\cal L}^R/i {\cal L}^K)^{3_F \bar 4_F} \qquad \rule[-6mm]{0mm}{0mm}
\\
\nonumber %    \label{eq:SigmaBF}
\left( \tilde \Sigma^{R/I}_{BF} \right)^{3_F \bar
 4_F}_{\bar 4_B 3_B} &= - 
%\left \{ \begin{array}{l} \!\!\! 1 \!\!\! \\ \!\!\! \toh \!\!\! 
%\end{array} \right\}
 \hbar^2 (\tilde G^{K/R})^{3_F \bar 4_F} \tilde G^A_{\bar 4_B 3_B}
 ({\cal L}^R/i {\cal L}^K)_{3_B}^{\;\;\;\;  \bar 4_F}
 \qquad \rule[-6mm]{0mm}{0mm}
\\
\nonumber %    \label{eq:SigmaFB}
 \left( \tilde \Sigma^{R/I}_{FB} \right)^{3_F \bar
  4_F}_{\bar 4_B 3_B} &= - 
%\left \{ \begin{array}{l} \!\!\! 1 \!\!\! \\ \!\!\! \toh \!\!\! 
%\end{array} \right\}
 \hbar^2 \tilde G^{R, 3_F \bar 4_F} (\tilde G^{K/A})_{\bar 4_B 3_B}
({\cal L}^R/i {\cal L}^K)^{3_F}_{\;\; \; \bar 4_B} 
 \qquad \rule[-6mm]{0mm}{0mm}
\\
 %    \label{eq:SigmaBB}
\left( \tilde \Sigma^{R/I}_{BB} \right)^{3_F \bar
  4_F}_{\bar 4_B 3_B} &=\! 
\phantom{-} 
%\left \{ \begin{array}{l} \!\!\! 1 \!\!\! \\ \!\!\! \toh
%  \!\!\! \end{array} \right\} \!
 \hbar^2 \tilde G^{R, 3_F \bar 4_F} 
(\tilde G^{K/A})_{\bar 4_B 3_B}
({\cal L}^R/i {\cal L}^K)_{3_B \bar 4_B} \; . 
 \qquad \rule[-1mm]{0mm}{0mm}
 \label{eq:selfenergies-explicit} 
     \end{align}
%\end{subequations}
%     where ${\cal L}^R = \tilde R$ and ${\cal L}^K = -2i \tilde I $
%     are the retarded and Keldysh parts of the interaction propagator.
%    These results for the  Cooperon self-energy are depicted
%     diagrammatically in      Fig.~\ref{fig:selfenergy}.


\begin{thebibliography}{99}
\bibitem{controversy} Most relevant references can be found in
  Ref.~\cite{GZS}, which  gives a useful overview of the controversy
  from GZ's point of view.
\bibitem{hbarto0} Approximation (iii) is [as] problematic [as (ii)], because it
  occurs \emph{inside} the argument of the density matrix $\bar \rho^a_0
  (\bmR, \bmP)$, which has an infinitely sharp step when  $T\to 0$.
% To   judge approximation (iii), we  note that 
  In the coordinate  representation, the density matrix can be written
  as  $\tilde \rho^0_\ij =  \left\{ \sum_{n=0}^\infty {\partial^n n_0
      (0) \over n!  }  \left[ {(- i \hbar \bnabla_{\bmr_i})^2 / 2 m +
        V_\imp (\bmr_i) - \mu \over 2 T }\right]^n \right\} \delta_\ij
  $. Now, GZ neglect the fact that $\hbar \bnabla_{\bmr_i}$ does not
  commute with $V_\imp (\bmr_i)$, arguing  that this is justified here
  in the semiclassical limit $\hbar \to 0$.  We disagree, since in the
  above expression for $\tilde \rho^0_\ij$  every neglected power of
  $\hbar$ is accompanied by at least one power of $1/T$, \ie\ the
  neglected terms are of order $\hbar / (T  \tau_{\rm char})$, where
  $\tau_{\rm char}$ is a time scale characteristic of the disorder
  potential (rough estimates suggest   $\tau_{\rm char} \simeq
  \tauel$). Thus, they can be neglected only for $T \gtrsim \hbar /
  \tauel$.
\bibitem{consistencychecks} Strictly speaking, this argument is
  slightly premature: one next  needs to establish whether or not such
  $\delta n_0 \neq 0 $ contributions, when summed over an entire
  diffusive path, could average to zero.  While estimating this would
  be difficult, the main point to be made here, though, is simple: the
  fluctuations $\delta \bar S_R$ \emph{have} to be analyzed carefully,
  and cannot simply be thrown away by hand.  Note that  the conclusion
  $T \gtrsim \hbar /\tauel$  is consistent with the estimate in
  Ref.~\cite{hbarto0}, and with the observation that GZ's results for
  $\tauphi(T)$ agree with those of AAK only if
  $\omega^{\textrm{AAK}}_{\max} (= T/\hbar) \gtrsim
  \omega^\GZ_{\max}(= 1/\tauel)$.
\bibitem{othercorrections}
By taking  the irreducible self energy  to be given
only by the diagrams of Fig.~\ref{fig:selfenergy}, one 
neglects diagrams with crossed or overlapping interaction lines.
However, these are smaller than those considered by
at least one power of $1/g$, where $g$ is the dimensionless
conductance.
\bibitem{Hartree}
Here %\Eq{eq:C1122afterderivatives},
we have dropped a  Hartree-like contribution 
$\tilde {\cal C}_{[11',22']}^\Hartree $ to
$\tilde {\cal C}_{[11',22']}$ 
which corresponds to
(II.47) and is dropped by GZ, too.
It vanishes in 
absence of interactions and corresponds to a renormalization of the
conductance which does not contribute to decoherence.
\bibitem{asymmetric}  By evoking the standard time-slicing definition
  for path integrals,   it is possible to properly  define $\int {\cal
    D} (\bmR \bmP) $ as $  \prod_{a = F,B} \prod_{n=1}^{M-1} \Bigl(
  \int d \bmr^a_{n} \Bigr) \prod_{n=1}^M \Bigl( \int {d \bmp^a_n \over
    (2 \pi )^d} \Bigr) e^{i (\dots)}$,  etc. In doing so, we choose to
  use asymmetric conventions for $a = F$ or $B$: If $\bmr^a_{n-1}$ and
  $\bmr^a_n$ are  the end-of-interval coordinates for time slice $n$,
  and $\delta \bmr_n^a = \bmr^a_n - \bmr^a_{n-1}$ is the relative
  coordinate [which is integrated over in the Fourier transforms of
  \Eq{eq:define-bar-rho}], then we choose the ``center-of-mass''
  coordinate  as $\bmR_n^F = \bmr^F_{n-1}$ and $\bmR_n^B =
  \bmr^B_{n}$, respectively.
\bibitem{SRSIcannotcancel} GZ have argued that  since $\bsr $ and
  $\bsi$ are both real functionals if their arguments are real, ``$i
  \bsr$ can never cancel any contribution from $\bsi$'' [discussion
  before (III.22)].  However, this argument overlooks the fact that
  $\bsr$ and $\bsi$ are both functionals of the path integral
  variables $\bmR (t_3)$ and $\bmP(t_3)$,  and that the integration
  measure $e^{i (\bar S_0^F - \bar S_0^B)}$  in $\int {\cal D} (\bmR
  \bmP)$ is \emph{complex} (GZ's argument would be correct if the
  integration measure were real).
\end{thebibliography}
\end{document}